\documentclass[10pt,pra,twocolumn,amssymb,nofootinbib,showpacs]{revtex4}
\usepackage{bm}
\usepackage{graphicx}
\usepackage{amsmath}
\begin{document}
\title{\bf Towards Quantum Superpositions of a Mirror:  an Exact Open Systems
Analysis -- Calculational Details}\altaffiliation{Send
correspondence to: \\ Stephen L. Adler \\ Institute for Advanced
Study \\ Einstein Drive, Princeton, NJ 08540 USA. \\ Phone:
609-734-8051 \\ Fax: 609-924-8399}

\author{Stephen L. Adler} \email{adler@ias.edu}
\affiliation{Institute for Advanced Study, Einstein Drive,
Princeton, NJ 08540, USA}

\author{Angelo Bassi}
\email{bassi@ictp.trieste.it} \affiliation{The Abdus Salam
International Centre for Theoretical Physics, Trieste \\
Istituto Nazionale di Fisica Nucleare, sezione di Trieste, Italy}

\author{Emiliano Ippoliti}\email{ippoliti@ts.infn.it}
\affiliation{Department of Theoretical Physics, University of
Trieste, Strada Costiera 11, 34014 Trieste, Italy  \\
Istituto Nazionale di Fisica Nucleare, sezione di
Trieste, Italy}

\begin{abstract}
We give details of calculations analyzing the proposed mirror
superposition experiment of Marshall, Simon, Penrose, and
Bouwmeester within different stochastic models for state vector
collapse. We give two methods for exactly calculating the fringe
visibility in these models, one proceeding directly from the
equation of motion for the expectation of the density matrix, and
the other proceeding from solving a linear stochastic unravelling
of this equation. We also give details of the calculation that
identifies the stochasticity parameter implied by the small
displacement Taylor expansion of the CSL model density matrix
equation. The implications of the two results are briefly
discussed. Two pedagogical appendices review mathematical
apparatus needed for the calculations.
\end{abstract}
 \maketitle

\section{Introduction}

There is currently much interest in experiments to create quantum
superposition states involving large numbers of particles, with
the ultimate aim of testing whether quantum superpositions of
macroscopic systems can be observed. Recently, Marshall, Simon,
Penrose, and Bouwmeester \cite{ref1}, motivated by suggestions of
Penrose \cite{ref2}, have proposed a novel interferometric
experiment in which a single photon interacts with a miniature
mirror mounted on a cantilever in one arm of the interferometer,
thus setting up a superposition of states containing of order
$10^{14}$ atoms. Since the two superposed states in this
experiment have a relative center of mass displacement of order
the width of the  mirror center of mass wave packet $\sim\sigma
\sim 10^{-11}$ cm, the experiment will place new constraints on
proposals for modifications to quantum mechanics in which center
of mass displacement is the key parameter.

Among the different proposals, collapse models
\cite{ref3,ref4,ref5,ref6} have been extensively studied. The
basic idea is to combine the standard Schr\"odinger  evolution and
the postulate of wavepacket reduction into one universal dynamical
equation, which is assumed to govern all physical processes. Such
a dynamics accounts both for the quantum properties of microscopic
systems and for the classical properties of macroscopic ones; in
particular, it guarantees that measurements made on microscopic
systems always have definite outcomes, and with the correct
quantum probabilities (Born probability rule).

In a recent Letter \cite{ref7}, we have analyzed the Marshall {\it
et al} experiment within the framework of the GRW \cite{ref3}, CSL
\cite{ref4} and QMUPL \cite{ref5} models, and have shown that
--- within the CLS model, which predicts the largest deviation
from standard quantum predictions --- one expects the maintenance
of coherence to better than 1 part in $10^8$.  Our aim in this
paper is to give the derivations of formulas presented, without
derivation, in our Letter. In Section 2 we give the basic
equations for the Marshall {\it et al} experiment, first as
formulated in their paper, and then as formulated within the
collapse models.  In Section 3 we solve for the visibility (the
physical quantity measured in the experiment) by direct
calculation from the density matrix evolution equation, making use
of the interaction picture, the Baker-Hausdorff formula, and
cyclic permutation under a trace. In Section 4 we give an
alternative derivation of the visibility, obtained by solving a
linear stochastic unravelling of the density matrix equation,
using the It\^o stochastic calculus. In Section 5 we compute the
stochasticity parameter entering into the visibility formula in
terms of the parameters of the CSL model. We briefly summarize our
results and their application to the Marshall {\it et al}
experiment in Section 6. In Appendix A we derive the
Baker--Hausdorff formula used in the text, and in Appendix B we
review the It\^o calculus formulas used in the calculation of
Section 4.

\section{Basic formalism}

The Hamiltonian for the Marshall {\it et al} experiment, with the
moving mirror in a cavity in interferometer arm $A$, is
\cite{ref9}
\begin{equation} \label{1}
H \; = \; \hbar \omega_{c}( a^{\dagger}_{A}a^{\phantom
\dagger}_{A} \, + \, a^{\dagger}_{B}a^{\phantom \dagger}_{B}) \; +
\; \hbar \omega_{m} b^{\dagger}b \; - \; \hbar G
a^{\dagger}_{A}a^{\phantom \dagger}_{A}(b + b^{\dagger}).
\end{equation}
Here $\omega_{c}$ is the frequency of the photon,
$a^{\dagger}_{A}$ and $a^{\dagger}_{B}$ are the creation operators
for the photon in the interferometer arms $A$ and $B$,
respectively, while $\omega_{m}$ and $b^{\dagger}$ are the
frequency and the phonon creation operator associated with motion
of the center of mass of the mirror. The coupling constant is
$G=\omega_{c} \sigma/L$, where $L$ is the length of the cavity,
with $\sigma=(\hbar/2M \omega_m)^{1\over 2}$ the width of the
mirror wave packet and  $M$ the mass of the mirror.

The semi--silvered beam splitter of the interferometer places the
photon in an initial state that is an equal superposition of being
in arm $A$ or $B$,
\begin{equation} \label{2}
|\psi_{0}\rangle \; = \;{1\over \sqrt{2}}\, \left[ |0\rangle_{A}
|1\rangle_{B} \; + \; |1\rangle_{A} |0\rangle_{B} \right]\,
|0\rangle_{m},
\end{equation}
and standard quantum mechanics predicts that at time $t$ the state
vector will be
\begin{eqnarray}\label{3}
|\psi_{t}\rangle & = & e^{-{i \over \hbar} H t} |\psi_{0}\rangle
\; = \; {1\over \sqrt{2}}e^{-i \omega_c t} \, \left[ \frac{}{}
|0\rangle_{A} |1\rangle_{B}|0\rangle_{m} \right. \nonumber \\
& + & \left. e^{i \kappa^2(\omega_m t-\sin \omega_m t)} \;
|1\rangle_{A} |0\rangle_{B}|\alpha_t \rangle_{m} \right].
\end{eqnarray}
Here we have written $\kappa =G/\omega_m$ and  $|\alpha_t
\rangle_{m} $ denotes a unit normalized mirror coherent state with
complex amplitude $\alpha_t =\kappa (1-e^{-i\omega_m t})$. While
in state $|0\rangle_{m}$ the mirror is fixed in its equilibrium
position (the origin of the reference frame), in state $|\alpha_t
\rangle_{m}$ the mirror oscillates between $0$ and $\ell \equiv
4\kappa\sigma$; in both cases, the shape of the wavefunction (in
position) is a Gaussian of width $\sigma$.

The physically measurable quantity considered by Marshall {\it et
al} is the maximum interference visibility for the photon
$\nu(t)$, defined as twice the modulus of the off--diagonal
element of the reduced density matrix of the photon. The full
density matrix for the system is
\begin{eqnarray} \label{4}
\rho & = & |\psi_{t}\rangle \langle \psi_{t}| \nonumber \\
& = & {1\over 2} \Big[ |0\rangle_A\, {}_A\langle 0| |1\rangle_B\,
{}_B\langle 1| |0\rangle_m\, {}_m\langle 0| \nonumber \\
& + & |1\rangle_A\, {}_A\langle 1| |0\rangle_B\, {}_B\langle 0|
|\alpha_t\rangle_m\, {}_m\langle \alpha_t| \nonumber \\
& + & e^{i\kappa^2(\omega_m t-\sin \omega_m t)} |1\rangle_A\,
{}_A\langle 0| |0\rangle_B\, {}_B\langle 1|  |\alpha_t\rangle_m\,
{}_m\langle 0| \nonumber \\
& + &  e^{-i\kappa^2(\omega_m t-\sin \omega_m t) } |0\rangle_A\,
{}_A\langle 1| |1\rangle_B\, {}_B\langle 0| |0\rangle_m\,
{}_m\langle \alpha_t| \Big]. \quad
\end{eqnarray}
Thus after tracing over the mirror states, the reduced density
matrix has as the coefficient of the off--diagonal term
$|1\rangle_A\, {}_A\langle 0|  |0\rangle_B\, {}_B\langle 1| $ the
factor ${1\over 2}f$, with
\begin{equation} \label{5}
f \; = \; e^{i\kappa^2(\omega_m t-\sin \omega_m t)} {}_m\langle 0|
\alpha_{t} \rangle_m,
\end{equation}
which using ${}_m\langle 0| \alpha_{t} \rangle_m =e^{-{1\over 2}
|\alpha_{t}|^2}$ gives
\begin{equation} \label{6}
f=  e^{i\kappa^2(\omega_m t-\sin \omega_m t)} e^{-\kappa^2(1-\cos
\omega_m t)}.
\end{equation}
Thus, under standard quantum mechanical evolution of the state,
one has for the time dependence of the visibility
\begin{equation} \label{7}
\nu(t) \; = \; e^{-\kappa^2(1-\cos \omega_m t)}.
\end{equation}
According to the above formula, the visibility starts from its
maximal value 1; it then decreases, but after half a period of the
mirror's motion it increases again, reaching the maximal value
after one period $T = 2\pi/\omega_{m}$. The strategy to test the
macroscopic superposition of the mirror then goes as follows. One
measures the photon's visibility after one period $T$: if it is
close to 1, then no collapse of the mirror's wavefunction has
occurred; if on the contrary it is smaller than 1, a spontaneous
collapse process is present which reduces the superposition to one
of its two terms. Of course, one must keep control of all sources
of decoherence, which tend to lower the observed visibility.

We proceed now to reanalyze the experiment using the modified
Schr\"odinger evolution of the QMUPL model of wavefunction
collapse \cite{ref5}; this model is particularly useful since, as
we shall prove, it allows one to get an exact formula for the
visibility when a spontaneous collapse mechanism is present.
Moreover, this model corresponds to the leading term in the
small--displacement Taylor expansion of both the GRW and the CSL
models; such an expansion is particularly suitable to the present
case since, according to the parameters of the experiment, the
maximum displacement between the two superposed states of the
mirror is of order $10^{-11}$ cm, which is much smaller than the
typical distance of $10^{-5}$ cm required for quantum
superpositions to be destroyed, in the GRW and CSL models. Under
the QMUPL model, the state vector evolves as
\begin{eqnarray}\label{8}
d\,|\psi_{t}\rangle & = & \left[ -{i \over \hbar}\, H\, dt +
\sqrt{\eta}\, (q - \langle q \rangle_{t})\, dW_{t} \right. \nonumber \\
& - & \left. {\eta\over 2}\, (q - \langle q \rangle_{t})^2 dt
\right] |\psi_{t}\rangle,
\end{eqnarray}
where $H$ is given by Eq. (\ref{1}), and $\langle q \rangle_{t}
\equiv \langle \psi_{t} | q | \psi_{t} \rangle$ is the quantum
mechanical expectation of the position operator $q=\sigma(b +
b^{\dagger}) $ associated with the center of mass of the mirror.
The stochastic dynamics is governed by a standard Wiener processes
$W_{t}$, defined on a probability space $(\Omega, {\cal F}, {\bf
P})$. Using the rules of the It\^o calculus (see Appendix B), the
density matrix evolution corresponding to Eq.(\ref{8}) is
\begin{equation}\label{9}
d\hat\rho=-{i\over \hbar} [H,\hat\rho] dt -{1\over 2} \eta
[q,[q,\hat\rho]] dt+ \sqrt{\eta} [\hat \rho,[\hat \rho,q]] dW_t.
\end{equation}

Since to observe interference fringes experimentally requires
passing to an ensemble of identically prepared photons through the
apparatus, the relevant density matrix in the stochastic case is
the ensemble expectation $\rho=E[\hat \rho]$, which obeys the
ordinary differential equation
\begin{eqnarray}\label{10}
d\rho\over dt & = & -{i\over \hbar} [H,\rho] -{1\over 2} \eta
[q,[q,\rho]] \nonumber \\
& = & -{i\over \hbar} [H,\rho] -{1\over 2} \eta \sigma^2
[b+b^{\dagger},[b+b^{\dagger},\rho]].
\end{eqnarray}
Defining an off--diagonal density matrix $\rho_{OD}$ acting in the
mirror Hilbert subspace by \hfill\break $ {}_A\langle 1|{}_B
\langle 0|  \rho |0\rangle_A|1\rangle_B ={1\over 2}\rho_{OD}$, so
that the factor $f$ introduced above is ${\rm Tr}_{m} \rho_{OD}$,
we can project out from Eq. (\ref{10}) the evolution equation for
$\rho_{OD}$,
\begin{eqnarray}\label{11}
d \rho_{OD}(t) \over dt & = & -iH^A\rho_{OD}(t) +i\rho_{OD}(t)H^B
\nonumber \\
& & -{1\over 2} \eta \sigma^2
[b+b^{\dagger},[b+b^{\dagger},\rho_{OD}(t)]],
\end{eqnarray}
with $\hbar H^{A}$ the effective mirror Hamiltonian acting when
the photon passes through interferometer arm A, and with $\hbar
H^{B}$ the corresponding effective mirror Hamiltonian acting when
the photon passes through arm B,
\begin{equation}\label{12}
H^A \; = \; \omega_mb^{\dagger}b-G(b+b^{\dagger}) \qquad H^B \; =
\; \omega_mb^{\dagger}b.
\end{equation}

We must now solve the dynamics represented by Eqs. (\ref{11}) and
(\ref{12}), or equivalently by Eq. (\ref{10}), so as to calculate
${\rm Tr}_{m} \rho_{OD}$ and obtain the visibility. Additionally,
we must calculate the stochasticity parameter $\eta$ entering into
Eqs. (\ref{8})--(\ref{11}) in terms of the parameters of the CSL
model. These are the issues addressed in the following three
sections.

\section{Direct solution for the visibility from the density matrix
evolution equation}

In this section we give a calculation of the mean visibility
directly from the density matrix equation of motion.  An essential
identity in everything that follows is the Baker-Hausdorff
identity, derived in Appendix A,
\begin{equation} \label{13}
e^{-iH^At}e^{iH^Bt} \; = \; N_{t}\, e^{\alpha_tb^{\dagger}} e^{
\beta_t b},
\end{equation}
with $H^A$ and $H^B$ as given in Eq. (\ref{12}), and with
\begin{eqnarray} \label{14}
N_{t} & = & e^{-\kappa^2(1-i\omega_mt - e^{-i\omega_m t} )}
\nonumber \\
& = & e^{-\kappa^2(1-\cos \omega_m t)+i\kappa^2(\omega_m t-\sin
\omega_m
t)} \nonumber \\
\alpha_t & = &\kappa(1-e^{-i\omega_m t} ) \\
\beta_t & = &-\kappa(1-e^{i\omega_m t}). \nonumber
\end{eqnarray}
Defining the photon off-diagonal part of the density matrix
$\rho_{OD}$ as in Section 2, which obeys the evolution equation of
Eq. (\ref{11}), the visibility is $\nu= |{\rm Tr}_m \rho_{OD}|$;
thus what is needed is to calculate  ${\rm Tr}_m \rho_{OD}$.

Let us now go to the interaction picture by defining
\begin{equation} \label{15a}
\rho_{OD}^I(t)=e^{iH^At} \rho_{OD}(t) e^{-i H^B t},
\end{equation}
so that $\rho_{OD}^I(0)=\rho_{OD}(0)=|0\rangle_m \, {}_m\langle
0|$. The corresponding differential equation obeyed by
$\rho_{OD}^I$ is
\begin{equation} \label{15b}
{d \rho_{OD}^I(t) \over dt} = -{1\over 2} \eta \sigma^2 e^{iH^A t}
[b+b^{\dagger},[b+b^{\dagger},\rho_{OD}(t)]]   e^{-i H^B t}.
\end{equation}
Multiplying from the left by $ e^{-iH^A u}$ and from the right by
$e^{i H^B u}$, we get the differential equation
\begin{eqnarray} \label{16}
\lefteqn{e^{-iH^Au} {d \rho_{OD}^I(t) \over dt}  e^{i H^B u} = }
\\ & & -{1\over 2} \eta \sigma^2  e^{iH^A(t-u)}
[b+b^{\dagger},[b+b^{\dagger},\rho_{OD}(t)]] e^{-i
H^B(t-u)}.\nonumber
\end{eqnarray}
Now take ${\rm Tr}_m$ of this equation, and use cyclic invariance,
to get
\begin{eqnarray} \label{17}
\lefteqn{ {d \over dt} {\rm Tr}_m e^{-iH^Au} \rho_{OD}^I(t)  e^{i
H^Bu} = } \\
& & \!\!\!\!\!\!\!-{\eta \sigma^2\over 2}   {\rm Tr}_m \!\!\left\{
[b+b^{\dagger},[b+b^{\dagger}, e^{-i H^B (t-u)} e^{iH^A(t-u)} ]]
\rho_{OD}(t)\right\}\!. \nonumber
\end{eqnarray}
Taking the adjoint of Eq. (\ref{13}) and setting $t\to -t$, we get
\begin{equation} \label{18}
e^{iH^Bt}  e^{-iH^At} \; = \; N_{t}\,e^{\beta_t b^{\dagger}} e^{
\alpha_t b},
\end{equation}
from which we easily calculate that the double commutator in Eq.
(\ref{17}) is
\begin{eqnarray} \label{19}
\lefteqn{ [b+b^{\dagger},[b+b^{\dagger}, e^{-i H^B (t-u)}
e^{iH^A(t-u)} ]]=} \\
& = & (\beta_{u-t}-\alpha_{u-t})^2  e^{-i H^B (t-u)} e^{iH^A(t-u)}
\nonumber \\
& = & 4\kappa^2\big(1-\cos\omega_m(u-t)\big)^2 e^{-i H^B (t-u)}
e^{iH^A(t-u)}. \nonumber
\end{eqnarray}

Substituting this into Eq. (\ref{17}), and using cyclic invariance
of the trace and Eq. (\ref{15a}), then gives
\begin{eqnarray} \label{20}
\lefteqn{ {d \over dt} {\rm Tr}_m e^{-iH^Au} \rho_{OD}^I(t)  e^{i
H^Bu}} \\
& = & -2 \eta (\kappa\sigma)^2 \big(1-\cos\omega_m(u-t)\big)^2
\nonumber \\
& & {\rm Tr}_m e^{-iH^Au} \rho_{OD}^I(t)  e^{i H^Bu}, \nonumber
\end{eqnarray}
which can be immediately integrated to give
\begin{eqnarray} \label{21}
\lefteqn{ {\rm Tr}_m e^{-iH^Au} \rho_{OD}^I(t)  e^{i H^Bu}} \\
& = & e^{ -2 \eta (\kappa\sigma)^2 \int_0^t dv
\big(1-\cos\omega_m(u-v)\big)^2 } \nonumber \\
& & {\rm Tr}_m e^{-iH^Au} \rho_{OD}^I(0)  e^{i H^Bu}. \nonumber
\end{eqnarray}
Setting $u=t$ in this equation, and using Eq. (\ref{15a}) and
cyclic invariance of the trace together with Eq. (\ref{18}), we
get
\begin{eqnarray} \label{22}
f & = & {\rm Tr}_m \rho_{OD}(t) \nonumber \\
& = & e^{ -2 \eta (\kappa\sigma)^2 \int_0^t
dv\big(1-\cos\omega_m(t-v)\big)^2  } \nonumber \\
& & {\rm Tr}_m e^{-iH^At} \rho_{OD}^I(0)  e^{i H^Bt} \nonumber \\
& = & e^{ -2 \eta (\kappa\sigma)^2 \int_0^t dv (1-\cos \omega_m
v)^2 } {}_m\langle 0| Ne^{\beta_t b^{\dagger}} e^{ \alpha_t b}
|0\rangle_m \nonumber \\
& = & e^{-{3\over 16} \eta \ell^2 (t-{4\over 3} {\sin \omega_m t
\over \omega_m } + {\sin 2 \omega_m t \over 6 \omega_m} ) }
\nonumber \\
& & e^{-\kappa^2(1-\cos \omega_m t)+i\kappa^2(\omega_m t-\sin
\omega_m t)}.
\end{eqnarray}
Finally, taking the absolute value of Eq. (\ref{22}), we get for
the visibility
\begin{eqnarray}\label{23}
\nu(t) & = & \exp\left[-\kappa^2(1-\cos \omega_m t)\right]
\\
& & \times \exp\left[-{3\over 16} \eta \ell^2 \left(t-{4\over 3}
{\sin \omega_m t \over \omega_m} + {\sin 2 \omega_m t \over 6
\omega_m}\right) \right]. \nonumber
\end{eqnarray}
Equations (\ref{22}) and (\ref{23}) are the results that we quoted
in Ref. \cite{ref7}.

\section{Solution for the visibility by a stochastic
unravelling method}

In this section we give an alternate derivation of Eq. (\ref{23}),
using stochastic methods to solve Eq. (\ref{10}).  We exploit the
property that although Eq. (\ref{8}) for the stochastic evolution
of the state vector uniquely implies the evolution of equation
(\ref{10}) for the expectation density matrix $\rho$, this
relationship is not one to one: there are in fact an infinite
number of different stochastic evolutions (or unravellings) which
imply Eq. (\ref{10}) for the evolution of their expectations
\cite{ref8}. In particular, a simple calculation using the It\^o
calculus shows that the {\it linear} stochastic equation
\begin{equation} \label{24}
d\,|\psi_{t}\rangle \; = \; \left[ -{i\over \hbar}\, H\, dt +
i\sqrt{\eta}\, q\, dW_{t} - {\eta\over 2}\, q^2 dt \right]
|\psi_{t}\rangle
\end{equation}
also has Eq. (\ref{10}) for the evolution equation for
$\rho=E[|\psi_{t}\rangle \langle \psi_{t} |]$. This means that, as
long as one is interested only in the {\it statistical} properties
of the system --- i.e. expectation values like ${\rm Tr}_m
\rho_{OD}(t)$ and the visibility --- one can choose freely to work
either with the stochastic evolution of Eq. (\ref{8}) or with the
stochastic evolution of Eq. (\ref{24}). Of course, for {\it
individual} realizations of the stochastic process, the two
equations Eq. (\ref{8}) and Eq. (\ref{24}) imply radically
different dynamics; in particular, Eq. (\ref{8}) induces the
collapse of the wavefunction, while Eq. (\ref{24}) does not.
However, for all physical quantities that depend only on the
expectation of the density matrix, the two evolutions give the
same answer.

Let us then  resort to Eq. (\ref{24}), since it is linear.
According to this equation, the initial state (\ref{2}) evolves as
follows:
\begin{equation} \label{25}
|\psi_{t}\rangle \; = \; {1\over\sqrt{2}}\, e^{-i \omega_{c} t}
\left[ |0\rangle_{A} |1\rangle_{B} |\phi^{0}_{t}\rangle_{m} \; +
\; |0\rangle_{A} |1\rangle_{B} |\phi^{1}_{t}\rangle_{m} \right],
\end{equation}
where the state vectors $|\phi^{0}_{t}\rangle_{m}$ and
$|\phi^{1}_{t}\rangle_{m}$ satisfy the following stochastic
differential equation for the mirror center of mass\footnote{We
have rewritten the Hamiltonian of Eq.~(1) in terms of the mirror
center of mass coordinate $x$, by reexpressing $b$ and
$b^{\dagger}$ in terms of $x$ and $-i\hbar d/dx$.}:
\begin{eqnarray} \label{26}
d\, \phi^{n}_{t}(x) & = & \left[ {i\hbar\over 2M}\, {d^2\over
dx^2}\, dt \, - \, {iM\omega_{m}^2\over 2\hbar}\, x^2\, dt \, + \,
i n g x\, dt \,  \right. \nonumber \\
& & \left. +\, i \sqrt{\eta}\, x\, dW_{t} - {\eta\over 2}\, x^2 dt
\right] \phi^{n}_{t}(x),
\end{eqnarray}
with $n = 0,1$, and with the coupling constant $g=G/\sigma$. We
now have to find the solution for the initial condition
$|\phi^{n}_{0}\rangle_{m} = |0\rangle_{m}$.

We take as a trial solution,
\begin{equation} \label{27}
\phi^{n}_{t}(x) \; = \left({M\omega_m\over \pi
\hbar}\right)^{1\over 4} \exp\left[ -a^{n}_{t}\, x^{2} \, + \,
b^{n}_{t}\, x \, + \, c^{n}_{t} \right],
\end{equation}
and by substituting it into Eq. (\ref{26}) and using the rules of
the It\^o calculus, we get the following set of equations for the
parameters $a^{n}_{t}$, $b^{n}_{t}$ and $c^{n}_{t}$,
\begin{eqnarray} \label{28}
d\,a^{n}_{t} & = & - {2i\hbar\over M}\, \left(a_{t}^{n}\right)^{2}
dt \, + \, {iM\omega_{m}^{2}\over 2\hbar}dt \qquad  a^{n}_{0}  =
{M\omega_{m}\over2\hbar}, \nonumber \\
d\,b^{n}_{t} & = & \left[ ing \, - \, {2i\hbar \over M} b^{n}_{t}
a^{n}_{t} \right] dt \, + \, i \sqrt{\eta} dW_{t} \qquad b^{n}_{0}
=  0, \nonumber
\\
d\,c^{n}_{t} & = &  {i\hbar\over2M} \left[
\left(b_{t}^{n}\right)^{2} \, - \, 2a^{n}_{t} \right] dt \qquad
c^{n}_{0} = 0.
\end{eqnarray}
The first two equations can be easily integrated and one gets
\begin{eqnarray} \label{29}
a^{n}_{t} & = & {M\omega_{m}\over2\hbar}, \\
b^{n}_{t} & = & {ng\over\omega_{m}}\, \left[ 1 - e^{-i\omega_{m}t}
\right] \, + \, i \sqrt{\eta} \int_{0}^{t} e^{-i \omega_{m} (t -
s)} dW_{s}.\nonumber
\end{eqnarray}
The factor $f$ previously introduced can be written as:
\begin{equation} \label{30}
f \; = \; \int_{-\infty}^{+\infty} E \left[ \phi^{0}_{t}(x)^*\,
\phi^{1}_{t}(x) \right] dx .
\end{equation}
We reverse the two operations of computing the statistical average
$E[...]$ and of taking the partial trace; the integration over $x$
gives
\begin{equation} \label{31}
\int_{-\infty}^{+\infty}\phi^{0}_{t}(x)^* \, \phi^{1}_{t}(x) \, dx
\; = \; \exp \left[{ (b^{0*}_{t} + b^{1}_{t})^2\over 8\,a_t} \, +
\, c^{0\ast}_{t} + c^{1}_{t}\right].
\end{equation}

As the final step, we have to take the average of Eq. (\ref{31})
with respect to the noise. To this end, we compute the stochastic
differential of the exponent, obtaining after some algebra
\begin{eqnarray} \label{32}
\lefteqn{d \, \left[ { (b^{0*}_{t} + b^{1}_{t})^2\over 8\,a_t} \,
+ \, c^{0*}_{t} + c^{1}_{t}\right]} \\
& = & {i\hbar\over 2M\omega_{m}^{2}} g^2 \left[ 1 -
e^{\displaystyle -i\omega_{m}t} \right] \, dt \; + \;
{i\sqrt{\eta} \hbar g \over M\omega_{m} } z_{t} \,dt, \nonumber
\end{eqnarray}
where $z_{t}$ is the stochastic process given by the formula
\begin{equation} \label{33}
z_{t} \; = \; \int_{0}^{t} \sin \omega_{m} (t - s) \, dW_{s}.
\end{equation}
The form of the stochastic integral of Eq. (\ref{33}) is such that
$z_{t}$ is a Gaussian stochastic process with zero mean, while the
correlation function is
\begin{eqnarray} \label{34}
K(t,s) & = & E[z_t \,z_s] \\
& = & \int_{0}^{{\rm min}(t,s)} \sin \omega_{m} (t - u) \sin
\omega_{m} (s - u) \, du. \nonumber
\end{eqnarray}
Equation (\ref{32}) shows that, as expected, $f$ is the product of
a ``deterministic'' part $f_{D}$, which does not depend on the
noise $z_{t}$, and a ``stochastic'' part $f_{S}$ which depends on
the noise. The deterministic part gives the result of Eq.
(\ref{6}),
\begin{eqnarray} \label{35}
f_{D} & = & \exp \left[ {i\hbar\over 2M\omega_{m}^{2}} g^2
\int_{0}^{t}\left( 1 - e^{\displaystyle -i\omega_{m}s} \right)
ds \right]  \nonumber \\
& = & e^{i\kappa^2(\omega_m t-\sin \omega_m t)}
e^{-\kappa^2(1-\cos \omega_m t)}.
\end{eqnarray}
We now have to compute the stochastic part,
\begin{equation} \label{36}
f_{S} \; = \;  E\left[ \exp \left( i\, {\sqrt{\eta}\hbar g \over
M\omega_{m}} \int_{0}^{t} z_{s} \, ds \right) \right].
\end{equation}
One easily recognizes, in the above formula, the definition of the
characteristic functional $\Phi[k_{t}]$ of the Gaussian stochastic
process $z_{t}$, with $k_{t} = \sqrt{\eta} \hbar g/ M\omega_{m}$.
One then has,
\begin{eqnarray} \label{37}
\lefteqn{f_{S} =  \exp \left[ - {\eta\over 2} \left({\hbar g\over
M \omega_{m}}\right)^2 \int_{0}^{t} ds_{1} \int_{0}^{t} ds_{2} \,
K(s_{1}, s_{2}) \right]} \\
& = & \exp \left[ -{3\over 16}\, \eta \, \ell^{2} \, \left( t -
{4\over 3\omega_{m}}\sin \omega_{m}t + {1\over 6\omega_{m}} \sin
2\omega_{m}t \right) \right] \nonumber
\end{eqnarray}
with $\ell = 4 \kappa \sigma$ the maximum excursion of the mirror
center of mass in its oscillation. (A derivation of Eq. (\ref{37})
directly from the It\^o calculus is given in Appendix B.) The
final result for the visibility $\nu = |f|$ is thus
\begin{eqnarray}\label{38}
\nu(t) & = & \exp\left[-\kappa^2(1-\cos \omega_m t)\right]
\\
& & \times \exp\left[-{3\over 16} \eta \ell^2 \left(t-{4\over 3}
{\sin \omega_m t \over \omega_m} + {\sin 2 \omega_m t \over 6
\omega_m}\right) \right]. \nonumber
\end{eqnarray}
as also obtained by the method of Sec. 3.

\section{Calculation of the stochasticity parameter from the
CSL model}

In this section we calculate the stochasticity parameter $\eta$
appearing in Eq. (\ref{8}), in terms of parameters that appear in
the CSL model for state vector collapse, which applies to systems
of identical particles treated by a field-theoretic approach (for
a similar calculation  based on properties of the complementary
error function, see Ghirardi, Pearle, and Rimini \cite{ref4},
Appendix C). The relevant CSL equation, taken from Eqs. (8.23) and
(8.24) of the review of Bassi and Ghirardi \cite{ref6}, can be
written as
\begin{equation} \label{39}
{\partial \over \partial t} \langle {\bf Q}^{\prime}|\rho|{\bf
Q}^{\prime \prime} \rangle = -\Gamma( {\bf Q}^{\prime},{\bf
Q}^{\prime \prime} ) \langle {\bf Q}^{\prime}|\rho|{\bf Q}^{\prime
\prime} \rangle,
\end{equation}
where
\begin{equation} \label{40}
\Gamma( {\bf Q}^{\prime},{\bf Q}^{\prime \prime} ) = {1\over 2}
\gamma \int d^3x [F( {\bf Q}^{\prime} -{\bf x}) -  F( {\bf
Q}^{\prime\prime} -{\bf x})]^2,
\end{equation}
and where
\begin{equation} \label{41}
F({\bf z})= \int d^3y D({\bf y}) \left( {\alpha \over 2 \pi}
\right)^{3\over 2} e^{-(\alpha /2)({\bf z} + {\bf y})^2}.
\end{equation}
Letting ${\bf d}= {\bf Q}^{\prime}-{\bf Q}^{\prime \prime}$ and
using translation invariance and space inversion symmetry, we can
rewrite Eq. (\ref{40}) as
\begin{equation} \label{42}
\Gamma( {\bf Q}^{\prime},{\bf Q}^{\prime \prime} ) = {1\over 2}
\gamma \int d^3x [F( {\bf x}+{\bf d}) -  F({\bf x})]^2,
\end{equation}
so that Taylor expansion gives for the leading small displacement term
(with summation on $i,j$ understood)
\begin{equation} \label{43}
\Gamma( {\bf Q}^{\prime},{\bf Q}^{\prime \prime} ) \simeq {1\over
2} \gamma \int d^3x d_id_j {\partial \over \partial x_i} F( {\bf
x}){\partial \over \partial x_j} F( {\bf x}).
\end{equation}

We now use the fact that, acting on the exponential within the
integral of Eq. (\ref{41}), ${\partial \over \partial z_i}$ is
equivalent to ${\partial \over \partial y_i}$, which can be
integrated by parts to act on the density $D$.  Since for density
distributions with cubic or higher symmetry we expect the
coefficient of $d_id_j$ in Eq. (\ref{43}) to be proportional to
$\delta_{ij}$, we can extract this coefficient by replacing
$d_id_j$ by $\delta_{ij}{\bf d}^2/3$, giving
\begin{equation} \label{44}
\Gamma( {\bf Q}^{\prime},{\bf Q}^{\prime \prime} ) \simeq {1\over
2} \gamma  C {\bf d}^2,
\end{equation}
with the coefficient $C$ given by
\begin{eqnarray} \label{45}
C & = & {1\over 3} \left( {\alpha \over 2 \pi} \right)^{3} \int
d^3y \int d^3w \; \partial_i  D({\bf y})  \partial_i  D({\bf w})
\nonumber \\
& & \times \int d^3x e^{-(\alpha /2)[({\bf x} + {\bf y})^2 + ({\bf
x} + {\bf w})^2 ]}.
\end{eqnarray}
We can now complete the square in the exponent,
\begin{equation} \label{46}
({\bf x} + {\bf y})^2 + ({\bf x} + {\bf w})^2 =2  [{\bf x}
+{1\over 2} ({\bf y}+{\bf w})]^2 + {1\over 2} ({\bf y}-{\bf w})^2,
\end{equation}
which allows us to do the ${\bf x}$ integration, giving
\begin{equation} \label{47}
C = {1\over 24} \left( {\alpha \over  \pi}
\right)^{\frac{3}{2}}\!\! \int d^3y \int d^3w \partial_i  D({\bf
y})  \partial_i  D({\bf w}) e^{-(\alpha /4) ({\bf y}-{\bf w})^2 }.
\end{equation}
Let us now assume a cubical volume of uniform density $D_0$ and
side $S$, so that we can take
\begin{equation} \label{48}
D({\bf w})=D_0\prod_{i=1}^3 \theta(w_i+S/2)\theta(S/2-w_i).
\end{equation}
The three terms summed over $i$ in Eq. (\ref{47}) give equal
contributions, so we have
\begin{equation} \label{49}
C \; = \; {1\over 8}D_0^2 \left( {\alpha \over  \pi} \right)^{3/2}
I_{12} I_3,
\end{equation}
with
\begin{eqnarray} \label{50}
I_{12} & = & \int_{-S/2}^{S/2}...\int_{-S/2}^{S/2} dy_1dy_2
dw_1dw_2 \; e^{-(\alpha/4) (y_1-w_1)^2} \nonumber \\
& & e^{-(\alpha/4) (y_2-w_2)^2},
\end{eqnarray}
and with
\begin{eqnarray} \label{51}
I_3 & = & \int_{-\infty}^{\infty} dy_3 \int_{-\infty}^{\infty}
dw_3 [\delta(y_3+S/2)-\delta(S/2-y_3)] \nonumber \\
& & [\delta(w_3+S/2)-\delta(S/2-w_3)] \; e^{-(\alpha/4)
(y_3-w_3)^2}. \;
\end{eqnarray}
When $S^2 \alpha >>1$, we can use the fact that the exponentials are
sharply peaked to get the approximations
\begin{equation} \label{52}
I_{12} \; \simeq \; S^2 4 \pi/\alpha \quad\qquad I_3 \; \simeq  \;
2,
\end{equation}
giving
\begin{equation} \label{53}
C \; \simeq \; D_0^2 S^2   \left( {\alpha \over  \pi}
\right)^{1/2}.
\end{equation}
This identifies the parameter $\eta$ appearing as the coefficient
of the $[q,[q,\rho]]$ term in the density matrix equation of
motion (\ref{10}) as
\begin{equation} \label{54}
\eta \; = \; \gamma C \; = \; \gamma S^2 D_0^2 \left( {\alpha
\over \pi} \right)^{1/2},
\end{equation}
as used in Eq. (14) of Ref. \cite{ref7}.

As a consistency check, let us use Eq. (\ref{54}) to determine the
transition regime from quadratic growth of $\Gamma$ to linear
growth.  For $|{\bf d}| \alpha^{1/2} >>1$, we know (see Bassi and
Ghirardi \cite{ref6}, p. 326) that $\Gamma$ is given by the
formula $\Gamma=\gamma n_{\rm out} D_0$, with $n_{\rm out}$ the
number of nucleons in the displaced cube not lying in the original
cube, which is clearly (for a 3 axis displacement) given by $|{\bf
d}| S^2 D_0$.  So equating $  (1/2)\gamma S^2 D_0^2
(\alpha/\pi)^{1/2}|{\bf d}|^2= \gamma  |{\bf d}| S^2 D_0^2$, we
find that the transition from quadratic to linear growth occurs at
$ |{\bf d}|= 2 (\pi/\alpha)^{1/2}$, which is of order the width of
the Gaussians and so is reasonable.

\section{Discussion}

To summarize, we have given details of the calculation of the
stochastic reduction in the visibility implied by Eqs.
(\ref{8})--(\ref{10}), leading to the visibility formula of Eqs.
(\ref{23}) and (\ref{38}), as well as details of the calculation
of the stochasticity parameter $\eta$ implied by the CSL model,
leading to the formula of Eq. (\ref{54}). As already discussed, in
the absence of stochastic reduction, the visibility as given by
Eq. (\ref{7}) starts at 1 at time $t=0$, decreases as $t$
increases, and then returns to 1 at $t=2\pi/\omega_m$, at which
point the mirror has completed one period of its oscillation.  By
contrast, with stochasticity present, we learn from Eqs.
(\ref{23}) and (\ref{38}) that at time $t=2\pi/\omega_m$ the
mirror visibility is damped by a factor $e^{-\Lambda}$, with
\begin{equation} \label{55}
\Lambda \; = \; (3/16) \eta \ell^2 (2\pi/\omega_m).
\end{equation}
Combining this formula with Eq. (\ref{54}), in the CSL model we
get
\begin{equation} \label{56}
\Lambda \; = \; (3/16) \gamma S^2 D_0^2 \left( {\alpha \over \pi}
\right)^{1/2}~ \ell^2 (2\pi/\omega_m).
\end{equation}
As shown in Ref. \cite{ref7}, which gives a detailed discussion of
the physical context, for the parameter values appropriate to the
CSL model and the Marshall {\it et al} experiment, Eq. (\ref{56})
gives $\Lambda \sim 0.2 \times 10^{-8}$, indicating that according
to the CSL model, coherence is maintained to an accuracy of better
than one part in $10^{8}$. Thus the Marshall {\it et al}
experiment is orders of magnitude away from a capability of
testing spontaneous collapse models for state vector reduction.

\section*{Acknowledgements}

The work of S.A. was supported in part by the Department of Energy
under Grant \#DE--FG02--90ER40542. The work of A.B. and E. I. was
supported in part by the Istituto Nazionale di Fisica Nucleare. We
wish to thank Dik Bouwmeester, Marco Genovese and GianCarlo
Ghirardi for helpful conversations.

\appendix
\section{Baker--Hausdorff formulas}

We derive here the Baker--Hausdorff formula of Eqs.
(\ref{13})--(\ref{14}). Let us define the unitary evolution
operator
\begin{equation} \label{A1}
U \; = \; e^{-iH^At},
\end{equation}
and the corresponding interaction picture operator
\begin{equation} \label{A2}
U^I \; = \; e^{iH^Bt} U= e^{iH^Bt} e^{-iH^At}.
\end{equation}
The operator $U^I$ obeys the equation of motion
\begin{eqnarray} \label{A3}
{dU^I\over dt} & = & e^{iH^Bt}  i(H^B-H^A)e^{-iH^Bt}U^I \nonumber
\\
& = & e^{iH^Bt} iG(b+b^{\dagger})e^{-iH^Bt}U^I \nonumber \\
& = & [A(t) + B(t)] U^I,
\end{eqnarray}
where we have defined
\begin{eqnarray} \label{A4}
A(t) & = & iGe^{iH^Bt}b  e^{-iH^Bt}=iGe^{-i\omega_m t} b,
\nonumber \\
B(t) & = & iGe^{iH^Bt}b^{\dagger} e^{-iH^Bt}=iGe^{i\omega_m t}
b^{\dagger}.
\end{eqnarray}
These obey the commutators
\begin{eqnarray} \label{A5}
[A(s),A(t)] & = & [B(s),B(t)] \; = \; 0,\nonumber \\
{[A(s),B(t)]} & = & -G^2\exp[ -i\omega_m(s-t)],
\end{eqnarray}
all of which are $c$-numbers. Integrating Eq. (\ref{A3}) with
respect to $t$, and using $U^I(0)=1$, we get
\begin{equation} \label{A6}
U^I(t) \; = \; T\,\exp\left[\int_0^t
ds\left(A(s)+B(s)\right)\right],
\end{equation}
where $T$ orders later times to the left.

Consider now the operator $W$ defined as
\begin{equation} \label{A7}
W \; = \; \exp\left[\int_0^t ds B(s)\right] \exp\left[\int_0^t ds
A(s)\right],
\end{equation}
which obeys
\begin{eqnarray} \label{A8}
{dW \over dt} & = & \exp\left[\int_0^t ds B(s)\right][B(t) + A(t)]
\nonumber \\
& &\exp\left[\int_0^t ds A(s)\right] \nonumber \\
& = & \left\{ B(t) + \exp\left[\int_0^t ds B(s)\right]A(t) \right.
\nonumber \\
& & \left. \exp\left[-\int_0^t ds B(s)\right]\right\}  W.
\quad\qquad
\end{eqnarray}
Now for general $u$ we have
\begin{eqnarray} \label{A9}
\lefteqn{{d \over dt} \exp\left[\int_0^t ds B(s)\right]A(u)
\exp\left[-\int_0^t ds
B(s)\right]} \nonumber \\
& = & \exp\left[\int_0^t ds B(s)\right][B(t),A(u)]
\exp\left[-\int_0^t ds B(s)\right]
\nonumber \\
& = & [B(t),A(u)],
\end{eqnarray}
where we have used the fact that the commutator $[B(t),A(u)]$ is a
$c$-number. Integrating on $t$, this gives
\begin{eqnarray} \label{A10}
\lefteqn{\!\!\!\!\!\!\!\!\!\!\!\!\!\!\!\!\!\!\!\!\!
\!\!\!\!\!\!\!\exp\left[\int_0^t ds B(s)\right]A(u)
\exp\left[-\int_0^t ds B(s)\right]}
\nonumber \\
& = & A(u) + \int_0^t ds [B(s),A(u)],
\end{eqnarray}
and now setting $u=t$ we get
\begin{eqnarray} \label{A11}
\lefteqn{\!\!\!\!\!\!\!\!\!\!\!\!\!\!\!\!\!\!\!\!\!
\!\!\!\!\!\!\!\exp\left[\int_0^t ds B(s)\right]A(t)
\exp\left[-\int_0^t ds B(s)\right]}
\nonumber \\
& = & A(t) + \int_0^t ds [B(s),A(t)].
\end{eqnarray}
Comparing with Eq. (\ref{A8}), we have obtained
\begin{equation} \label{A12}
{dW \over dt} \; = \; \left(A(t)+B(t)+\int_0^t ds
[B(s),A(t)]\right) W,
\end{equation}
and comparing this with Eqs. (\ref{A3}) and (\ref{A6}) for $U^I$,
we get \cite{ref10}
\begin{equation} \label{A13}
U^I \; = \; W\exp\left(-\int_0^t du \int_0^u ds[B(s),A(u)]
\right).
\end{equation}
Multiplying Eq. (\ref{A2}) from the left by $e^{-iH^Bt}$  and from
the right by $e^{iH^Bt}$, we then get
\begin{widetext}
\begin{eqnarray} \label{A14}
e^{-iH^At} e^{iH^Bt} & = & e^{-iH^Bt} U^I e^{iH^Bt}  \nonumber
\\
& = & e^{-iH^Bt} \exp\left[\int_0^t ds B(s)\right] e^{iH^Bt}
e^{-iH^Bt}\exp\left[\int_0^t ds A(s)\right]  e^{iH^Bt}
\exp\left(-\int_0^t du
\int_0^u ds[B(s),A(u)] \right) \nonumber \\
& = & \exp\left[ e^{-iH^Bt} \int_0^t ds B(s) e^{iH^Bt}\right]
\exp\left[e^{-iH^Bt} \int_0^t ds A(s) e^{iH^Bt}\right]
\exp\left(-\int_0^t du \int_0^u
ds[B(s),A(u)] \right) \nonumber \\
& = & \exp\left[ \int_0^t ds\, iG e^{i\omega_ms} e^{-iH^Bt}
b^{\dagger} e^{iH^Bt}\right] \exp\left[\int_0^t ds\, iG
e^{-i\omega_ms} e^{-iH^Bt} b
e^{iH^Bt}\right] \nonumber \\
& \times & \exp\left(-\int_0^t du \int_0^u ds\,
G^2e^{i\omega_m(s-u)}
\right) \nonumber \\
& = & \exp\left[ \int_0^t ds\, iG e^{i\omega_ms} e^{-i\omega_mt}
b^{\dagger}\right] \exp\left[ \int_0^t ds\, iGe^{-i\omega_ms}
e^{i\omega_m t} b \right] \, \exp\left(-\int_0^t du \int_0^u ds\,
G^2e^{i\omega_m(s-u)}
\right) \nonumber \\
& = & \exp[\kappa (1- e^{-i\omega_mt}) b^{\dagger} ] \exp[ -\kappa
(1- e^{i\omega_m t} ) b]
\exp\left[-\kappa^2(1-i\omega_mt-e^{-i\omega_mt}) \right],
\end{eqnarray}
\end{widetext}
which is Eq. (\ref{14}) of Sec. 3.

\section{Basic It\^o calculus formulas}

The stochastic differential $dW_t$ behaves heuristically as a random
square root of $dt$, as expressed in the It\^o calculus rules
\begin{equation} \label{B1}
dW_t^2 \; = \; dt, \quad\qquad dW_t dt \; = \; dt^2=0.
\end{equation}
As a consequence of Eq. (\ref{B1}), the Leibniz chain rule of the
usual calculus is modified to
\begin{equation} \label{B2}
d(AB) \; = \; dA\; B + A \;dB + dA\; dB,
\end{equation}
and thus in differentiating a function $f(A)$, one has
\begin{equation} \label{B3}
df(A)=f(A+dA)-f(A) = f^{\prime}(A) dA + {1\over 2}
f^{\prime\prime}(A) (dA)^2.
\end{equation}
These formulas are used in the calculations leading to Eqs.
(\ref{28}) and (\ref{32}) of Sec. 4.

The It\^o differential $dW_t$ is statistically independent of the random
process up to time t, so we have the definition
\begin{equation} \label{B4}
E[dW_t C(t)] \; = \; 0
\end{equation}
for any stochastic process $C(t)$ constructed from $dW_s$ with $s
\leq t$. From Eqs. (\ref{B1})--(\ref{B4}), we get useful formulas
for expectations of integrals. Consider first
\begin{equation} \label{B5}
f(t) \; = \; E\left[\int_0^t dW_u A(u) \int_0^tdW_u  B(u)\right],
\end{equation}
which has the differential
\begin{eqnarray} \label{B6}
df(t) & = & E\left[ dW_t A(t) \int_0^tdW_u  B(u) \right. \nonumber \\
& & + \left. \left(\int_0^t dW_u A(u)\right) dW_t  B(t) + A(t)
B(t) dt \right]
\nonumber \\
& = & E[A(t) B(t)] dt,
\end{eqnarray}
which integrates back to give
\begin{equation} \label{B7}
E\left[\int_0^t dW_u A(u) \int_0^tdW_u  B(u) \right] = \int_0^t du
E[A(u) B(u)],
\end{equation}
a formula called the It\^o isometry. When $A(u)$ and $B(u)$  have
differing domains of support $D_A$ and $D_B$, the integral on the
right of Eq. (\ref{B7}) clearly extends only over the intersection
$D_A \cap D_B$. Applying Eq. (\ref{B7})  to the definition of
$z_t$ in Eq. (\ref{33}) immediately gives the formula for the
correlation function $K(t,s)$ of Eq. (\ref{34}). Consider next the
expectation
\begin{equation} \label{B8}
f(t) \; = \; E\left[\exp\left(\int_0^t \Phi(u,v) dW_v \right)
\right].
\end{equation}
Its differential is, by Eq. (\ref{B3}),
\begin{eqnarray} \label{B9}
df & = & E\left[\exp\left(\int_0^t \Phi(u,v) dW_v \right)\right.
\nonumber \\
& &  \left.\left(\Phi(u,t) dW_t + {1\over 2}  \Phi(u,t)^2
dt\right) \right]
\nonumber \\
& = & {1\over2}\,f(t) \Phi(u,t)^2 dt,
\end{eqnarray}
which integrates back to give
\begin{equation} \label{B10}
E\left[\exp\left(\int_0^t \Phi(u,v) dW_v \right) \right] =
\exp\left({1\over 2}\int_0^t dv \Phi(u,v)^2 \right).
\end{equation}
In particular, setting $u=t$ we get the useful
formula
\begin{equation} \label{B11}
E\left[\exp\left(\int_0^t \Phi(t,v) dW_v \right) \right] =
\exp\left({1\over 2}\int_0^t dv \Phi(t,v)^2 \right).
\end{equation}
As an application of Eq. (\ref{B11}), consider the expectation
$g(t) =E[\exp(C\int_0^t z_s ds)]$, with $z_t$ given by Eq.
(\ref{33}). Since
\begin{eqnarray} \label{B12}
\int_0^t z_s ds & = & \int_0^t ds \int_0^s \sin\omega_m(s-v)
dW_v \nonumber \\
& = & \int_0^t dW_v \int_v^t ds  \sin\omega_m(s-v) \nonumber \\
& = & \int_0^t \omega_m^{-1}[1-\cos\omega_m(t-v)] dW_v, \qquad
\end{eqnarray}
the expectation $g(t)$ has the form of Eq. (\ref{B11}), with
$\Phi(t,v)= C\omega_m^{-1}[1-\cos\omega_m(t-v)]$, and we have
\begin{equation} \label{B13}
g(t) \; = \; \exp\left({C^2\over 2\omega_m^2} \int_0^t
[1-\cos\omega_m(t-v)]^2 dv\right),
\end{equation}
which corresponds to the integral appearing in Eq. (\ref{22}). An
alternative expression for $g(t)$ is obtained by using the formula
$\Phi(t,v)= C\int_v^t ds \sin \omega_m(s-v)$, which gives
\begin{widetext}
\begin{eqnarray} \label{B14}
\int_0^t dv \Phi(t,v)^2 & = & C^2\int_0^t dv  \int_v^t ds_1
\int_v^t ds_2 \sin \omega_m(s_1-v) \sin \omega_m(s_2-v) \nonumber
\\
& =& C^2\int_0^t ds_1 \int_0^t ds_2 \int_0^{{\rm min}(s_1,s_2)} dv
\sin \omega_m(s_1-v) \sin \omega_m(s_2-v) \nonumber \\
& = & C^2\int_0^t ds_1 \int_0^t ds_2 K(s_1,s_2),
\end{eqnarray}
\end{widetext}
with $K$ the correlation function defined in Eq. (\ref{34}). When
substituted into Eq. (\ref{B11}), this corresponds to the integral
appearing in Eq. (\ref{37}).

\end{document}